\documentclass[showpacs,aps,nofootinbib]{revtex4-2}
\emergencystretch=\maxdimen
\usepackage{bm}
\usepackage[utf8]{inputenc}
\usepackage{graphicx, psfrag}
\usepackage{amssymb}
\usepackage[colorlinks=true, citecolor=blue, urlcolor = blue, linkcolor= red, bookmarks=true]{hyperref}
\usepackage{color}
\usepackage{amsmath}
\usepackage{tensor}
\usepackage{amsfonts}
\usepackage{dcolumn}
\usepackage{pgfplots}
\usepackage{epstopdf}
\usepackage{booktabs}

\def\qq{\qquad}

\def\lal{&&{}}
\def\eq{Eq.\,}
\def\eqs{Eqs.\,}
\def\beq{\begin{equation}}
\def\eeq{\end{equation}}
\def\bear{\begin{eqnarray}}
\def\bearr{\begin{eqnarray} \lal}
\def\ear{\end{eqnarray}}
\def\earn{\nonumber \end{eqnarray}}
\def\nn{\nonumber\\ {}}

\def\yyy{\\[5pt] \lal }

\def\d{\partial}

\def\sign{\mathop{\rm sign}\nolimits}
\def\diag{\mathop{\rm diag}\nolimits}

\def\const{{\rm const}}

\def\ep{\epsilon}
\def\then{\ \Rightarrow\ }

\def\eqn#1{\eq\eqref{#1}}
\def\rf{\eqref}
\def\mn{_{\mu\nu}}
\def\MN{^{\mu\nu}}
\def\mN{_\mu^\nu}

\def\R{{\mathbb R}}
\def\GR{general relativity}
\def\sph{spherically symmetric}
\def\ssph{static, spherically symmetric}
\def\bh{black hole}
\def\bhs{black holes}
\def\wh{wormhole}

\def\asflat{asymptotically flat}
\def\emag{electromagnetic}
\def\Scw{Schwarz\-schild}
\def\RN{Reiss\-ner-Nord\-str\"om}

\begin{document}

\title{Field sources for Simpson--Visser space-times}

\author{Kirill A. Bronnikov}\email{kb20@yandex.ru}
\affiliation{Center of Gravitation and Fundamental Metrology, VNIIMS, Ozyornaya St. 46, Moscow 119361, Russia}
\affiliation{Peoples' Friendship University of Russia, 6 Miklukho-Maklaya St, Moscow, 117198, Russia}
\affiliation{National Research Nuclear University "MEPhI", Kashirskoe sh. 31, Moscow 115409, Russia}

\author{Rahul Kumar Walia} \email{KumarR@ukzn.ac.za}
\affiliation{Astrophysics Research Centre, School of Mathematics, Statistics and Computer Science, University of
	KwaZulu-Natal, Private Bag 54001, Durban 4000, South Africa}

\date{\today}

\begin{abstract}
  Simpson--Visser (SV) space-times are the simplest globally regular modifications of the \Scw, \RN\ and 
  other \bh\ solutions of \GR. They smoothly interpolate between these black holes and traversable wormholes. 
  After a brief presentation of the \Scw-like and \RN-like SV geometries, including their Carter-Penrose diagrams,
  we show that any \ssph\ SV metric can be obtained as an exact solution to the Einstein field equations 
  sourced by a combination of a minimally coupled phantom scalar field with a nonzero potential $V(\phi)$ 
  and a magnetic field in the framework of nonlinear electrodynamics with the Lagrangian ${\cal L(F)}$, 
  ${\cal F} = F\mn F\MN$ (in standard notations). Explicit forms of $V(\phi$ and ${\cal L(F)}$ are presented 
  for the cases of \Scw-like and \RN-like SV metrics.
\end{abstract}	
\maketitle

\section{Introduction}

  According to general relativity, under regular initial conditions, a sufficiently massive matter distribution 
  must undergo a complete gravitational collapse, forming a zero proper volume end state with a 
  curvature singularity, from which even light cannot escape \cite{Penrose:1964wq, Hawking:1969sw}. 
  At the singularity, physical quantities such as the space-time curvature, tidal forces, energy density, 
  and pressure, tend to infinity thus indicating a pathology of the classical theory of gravity. Since a singularity 
  is an undesirable feature of the theory, there emerge strategies to conceal it from an observer. The 
  Cosmic Censorship Conjecture suggests that space-time singularities are necessarily enclosed within a 
  null hypersurface, causally disconnecting the interior from the exterior, known as horizons \cite{Penrose:1969pc}. 
  Such an end state is known as a black hole. However, under generalized initial conditions for varieties of 
  matter fields, a number of possible end states have been reported where singularities are observable from 
  null infinity, being called naked singularities \cite{Yodzis:1973gha,Eardley:1978tr, Joshi:1992vr, 
  Christodoulou:1984mz, joshi, Ori:1987hg}.
  
  Another way to avoid singularities while remaining in the framework of classical gravity is to find regular 
  \bh\ models with finite space-time curvature at the center (e.g., 
  \cite{Bartnik:1988am,Bardeen:1968,Hayward:2005gi,Bronnikov:2000yz,dym92,kb-dym12}, and many others).
  One of the most popular classes of such regular black holes are those sourced by nonlinear electrodynamics (NED), 
  see, e.g.,  \cite{AyonBeato:1998ub,Bronnikov:2000yz,kb01-NED,kb18-NED} and references therein. 
  Singularity-free regular end states without a horizon are also possible, for example, in conformally flat collapsing star 
  models of $f(R)$ gravity \cite{Chakrabarti:2018qsh} or with the nonminimally coupled Gauss-Bonnet 
  invariant \cite{Chakrabarti:2017apq}. Such soliton-like configurations also emerge with NED sources.
  
  Quite a different kind of models show pathways connecting distant parts of the same universe or two 
  different universes --- these are wormholes and certain classes of regular \bhs. Among the latter, there are  
  \bhs\ containing an expanding cosmology beyond a horizon, the so-called black universes 
  \cite{Bronnikov:2005gm,Bronnikov:2006fu}. Such models are naturally obtained if one considers local 
  concentrations of dark energy in the form of a phantom matter field as well as in some modified gravity
  theories \cite{Bronnikov:2012ch, Bronnikov:2018vbs, Bronnikov:2006fu, Clement:2009ai,Azreg-Ainou:2011gcq}. 
  Thus, globally regular \ssph\ solutions with a magnetic field, describing both wormholes with flat or anti-de Sitter
  asymptotics and black universes, were obtained \cite{Bolokhov:2012kn, Bronnikov:2015kea}. 
  
  A discussion of wormhole geometries can be traced back to the papers by Flamm \cite{Flamm} (1916) and 
  Einstein and Rosen \cite{ER}, though the ``tunnels'' they describe only exist in some spatial sections of 
  space-time and are not traversable. As became clear much later \cite{Thorne88, Hochberg97}, maintaining 
  a static wormhole throat needs an amount of exotic matter that violates the Null Energy Condition (NEC). 
  Ellis \cite{Ellis} and Bronnikov \cite{kb73} discussed such a static traversable wormhole on the basis of a
  solution of Einstein gravity coupled to a free phantom scalar field, and quite a lot of other \wh\ solutions
  with various kinds of phantom matter were obtained and discussed afterwards. One can note that although 
  phantom fields have always been a source of debate due to their negative kinetic energy, they are a strong 
  candidate for dark energy and lead to some unique and interesting space-times \cite{Alam:2003fg, Allen:2004cd}. 
  Furthermore, phantom fields also appear in string theory in the form of negative tension branes, which play an 
  important role in string dualities \cite{Hull:1998ym, Okuda:2006fb}. Interestingly, a minimally coupled scalar 
  field can smoothly pass on from canonical to phantom form without creating any space-time singularities 
  leading to the ``trapped ghost'' concept \cite{Kroger:2003qh, Bronnikov:2010hu, Bronnikov:2018vbs}.
  Quite recently, examples of \ssph\ \wh\ models were obtained using classical spinor field sources
  \cite{radu21, roman21a, roman21b, wald21, kb21-note}, which can thus acquire exotic properties.  

  Concerning \bhs, one simple approach to regularize a \ssph\ geometry with a curvature singularity at $r=0$ 
  is to simply replace $r$ with $\sqrt{x^2+a^2}$ (and $dr$ with $dx$) \cite{Simpson:2018tsi}, where $x$ is a new 
  radial coordinate, and $a$ is some positive constant. The result is a globally regular space-time where the 
  spherical (areal) radius $r$ always remains positive, while the singularity at $r=0$ turns into a regular minimum 
  of $r$, a sphere of radius $a$. Due to regularity at $x = 0$, the metric can be extended to $x < 0$. Depending \
  on the nature of the hypersurface $x=0$, the resulting metric describes different geometries of interest: 
  a timelike throat $x=0$ corresponds to a reflection-symmetric traversable wormhole, while a spacelike hypersurface 
  $x=0$ corresponds to a bounce in the time-dependent quantity $r$, which is in fact one of the scale factors 
  of a Kantowski-Sachs cosmological model inside a \bh. This phenomenon was named a {\it black bounce}
  \cite{Simpson:2018tsi}. (In fact, a black bounce is also a necessary feature of all black-universe models.) 
  The intermediate case, a null throat, corresponds to what can be called a one-way wormhole but actually 
  possesses \bh\ properties since there is a horizon. 
  
  Simpson and Visser (SV) \cite{Simpson:2018tsi} considered such a metric on the basis of the \Scw\ one, 
  and the result is a family of metrics that seamlessly interpolates between a Schwarzschild black hole, a 
  black-bounce geometry, and a traversable wormhole. The SV metric is interesting in the sense that it is a 
  minimal one-parameter extension of the Schwarzschild metric. In the same spirit, a black-bounce extension 
  of the Reissner-Nordstr\"{o}m space-time was also reported \cite{Franzin:2021vnj}. The resulting SV metric 
  generalizes the \Scw-SV spacetime and interpolates between a \RN\ black hole and a wormhole. 
  Lobo \textit{et al.} \cite{Lobo:2020ffi} used this approach to construct a large family of globally regular black 
  bounce space-times that generalize the original SV model. Owing to their rich features, SV space-times have 
  been a subject of great interest over the recent years, and their rotating extensions have also been obtained 
  \cite{Mazza:2021rgq, Xu:2021lff, Shaikh:2021yux}. Interestingly, the rotating SV metric belongs to a special 
  case of the parameterized non-Kerr metric constructed by Johannsen \cite{Johannsen:2013szh}. The 
  geometric properties of black hole/black bounce space-times are studied by using gravitational wave echo signals 
  \cite{Bronnikov:2000yz, Churilova:2019cyt, Yang:2021cvh}. Their optical appearance has been investigated, 
  including the effects of a surrounding accretion disk and strong gravitational lensing \cite{Guerrero:2021ues,
  Tsukamoto:2021caq,Islam:2021ful,Cheng:2021hoc,Bronnikov:2021liv,Tsukamoto:2020bjm,Nascimento:2020ime}.   
 
  It should be noted that Lagrangian formulations for SV-like black bounce space-times seem to be still lacking. 
  In this paper, we show that both uncharged and charged SV metrics as well as their possible modifications and   
  extensions can be obtained as exact solutions to Einstein's field equations minimally coupled with a self-interacting 
  phantom scalar field combined with a NED field. It is shown that a scalar field and NED taken separately cannot 
  create an SV black bounce space-time. The need for a phantom scalar field was expected due to the presence 
  of a wormhole as a solution to the field equations, while a NED ingredient accounts for adjusting the stress-energy 
  tensor to a form relevant to particular SV space-times. 

  The paper is organized as follows. In Section~\ref{sec-2} we discuss the general properties of uncharged and 
  charged SV spacetimes, including their global features characterized by Carter-Penrose diagrams. 
  In Section~\ref{sec-3} we present the scalar-NED fields model for SV spacetimes. 
  Our concluding remarks are contained in Section~\ref{sec-5}. We adopt the metric signature $(+,\,-\,,-\,,-)$ 
  and work in the geometrized units $8\pi G = c = 1$.
		
\section{Simpson--Visser space-times}\label{sec-2}

\begin{figure*}
\centering
\includegraphics[scale=0.85]{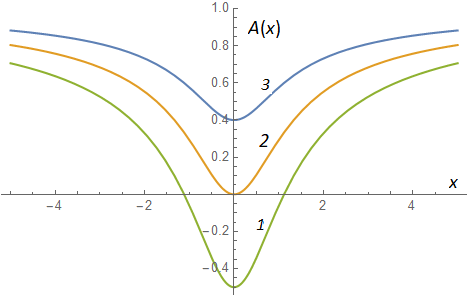}
\includegraphics[scale=0.85]{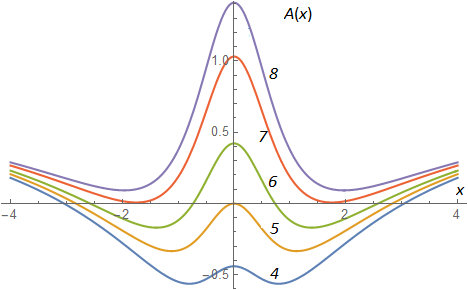}
\caption{\small
	The metric function $A(x)$ of the SV metric \rf{ds-gen}, with $a=1$ taken as the length scale. 
	Left panel: $A(x) = 1 - 2M/r(x)$, curves 1--3 correspond to $M = 0.75,\ 0.5,\ 0.3$, respectively.
	Right panel: $A(x) = 1 - 2M/r(x) + Q^2/r^2(x)$, curves 4--8 correspond to $M = 2$
	and $Q= 1.6,\ 1.732,\ 1.85,\ 2.008,\ 2.1$, respectively.}
	\label{f1}
\end{figure*}

  The Simpson-Visser (SV) space-time metric \cite{Simpson:2018tsi} is a one-parameter modification of the 
  Schwarzschild metric (to be called the \Scw-SV, or S-SV metric), 
  in which the spherical radius $r$ is replaced by the expression $\sqrt{x^2 + a^2}$, where $x \in \R$ 
  is a new radial coordinate, and $a = \const > 0$ is a new parameter, so that the metric has the form
\beq        \label{ds-gen}
		ds^2 = A(x) dt^2 - \frac {dx^2}{A(x)} - r^2(x) (d\theta^2+\sin^2\theta\,d\varphi^2),
		\qq\     r(x) \equiv \sqrt{x^2 + a^2},
\eeq  
  with $A(x) = 1 - 2M/r(x)$, and $M >0$ has the meaning of the \Scw\ mass. Unlike the \Scw\ metric
  (restored in the case $a=0$), at $a > 0$ this metric is manifestly globally regular and twice (as $x\to\pm\infty$)
  \asflat. The new parameter $a$ controls interpolation between the Schwarzschild \bh\ metric and that of a
  a Morris-Thorne-like traversable wormhole, see Fig.\,\ref{f1}, left panel. Thus we obtain the following 
  space-time geometries:
\begin{itemize}
\item 
	$2M > a$ --- a regular black hole with two horizons at $x_h^{\pm}=\pm \sqrt{4M^2-a^2}$ (curve 1
	in Fig.\,1).
\item 
	$2M = a$ --- a regular black hole with a single extremal (double) horizon at $x=0$ (curve 2).
\item 
	$2M < a $ --- a symmetric traversable wormhole with a throat at $x=0$ (curve 3). In particular, 
	at $M=0$ we obtain the Ellis \wh, free from static gravitational forces as well as tidal forces \cite{Ellis, kb73} 
\end{itemize}  

\begin{figure*}
\centering
\includegraphics[width=16cm]{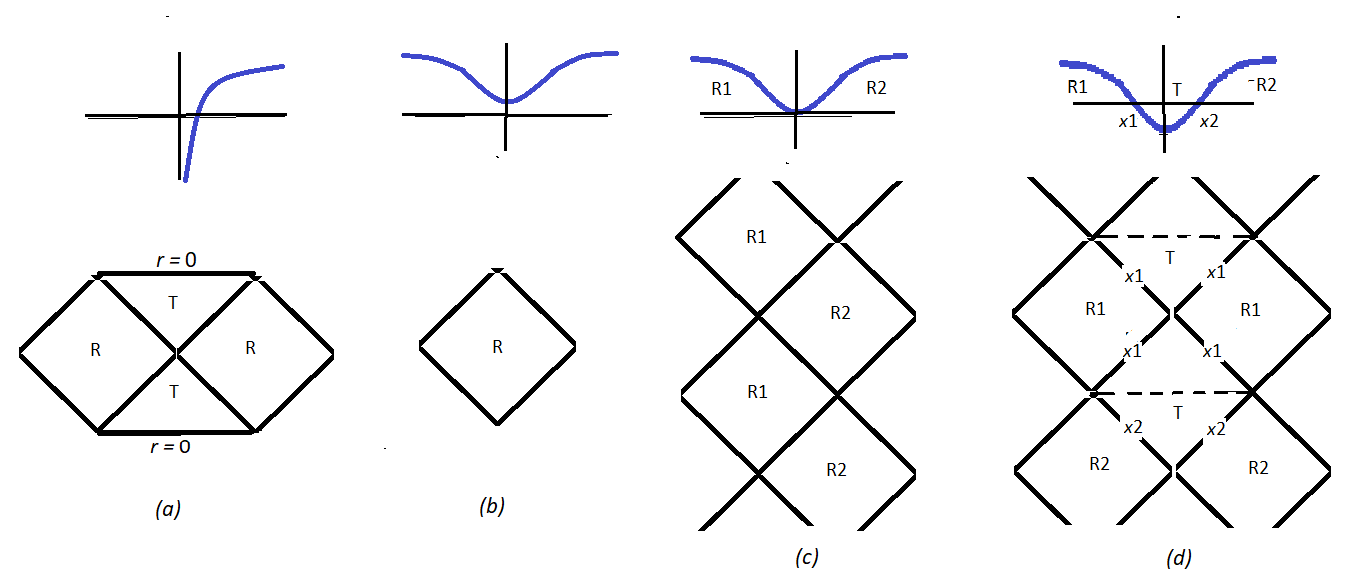}
\caption{\small
	Carter-Penrose diagrams (a) for the \Scw\ geometry (for comparison) and (b,c,d) for S-SV
	geometries. Over the diagrams, the corresponding behavior of $A(x)$ is schematically shown. 
	All tilted boundary lines in the diagrams correspond to $r = \infty$, all tilted internal lines depict horizons.
	The letters R and T mark R- and T-regions, enumerated where necessary, $x1$ and $x2$ mark
	the positions of horizons. The black bounce at $x=0$ is shown by dashed lines in the T-region.
	The diagrams (c) and (d) are infinitely continued upward and downward.
	}
		\label{f2}
\end{figure*}

  Thus the SV geometries are not only globally regular but are also richer in their causal structure than the 
  \Scw\ geometry, as can be illustrated by their Carter-Penrose diagrams, see Fig.\,\ref{f2}. A regular minimum 
  of the spherical radius $r(x)$ occurring in a nonstatic ($A(x) < 0$) space-time region, also often called a 
  T-region, was named a ``black bounce'' \cite{Simpson:2018tsi}. One can note that geometries containing 
  a black bounce, described by solutions to the Einstein equations with phantom scalar fields, were 
  considered earlier, in particular, in \cite{Bronnikov:2005gm,Bronnikov:2006fu,Bronnikov:2012ch,
  Bolokhov:2012kn,Bronnikov:2015kea,Bronnikov:2018vbs}, 
  and in many cases led to so-called ``black universes'' \cite{Bronnikov:2005gm,Bronnikov:2006fu},
  i.e., geometries where the T-region of a \bh\ ultimately evolves into a de Sitter-like expanding universe.
  
  In the case $2M =a$, such that the minimum of $r$ coincides with a single horizon separating two R-regions,
  the geometry has also been called ``a one-way wormhole'' \cite{Simpson:2018tsi}, but it has definitely 
  a \bh\ nature and looks from outside quite similarly to an extremal \RN\ \bh. A proper name for such 
  a horizon may be a ``black throat.'' 
   

  Later on, Franzin \textit{et al.} \cite{Franzin:2021vnj} considered more general black-bounce space-times 
  on the basis of the Reissner--Nordstr\"{o}m metric in the same minimal manner, leading to a number of
  geometries including regular black holes and traversable wormholes. The RN-SV metric has the form
  \rf{ds-gen} with 
\beq                   \label{RN-SV}
		A(x) = 1 - \frac{2 M}{r(x)} + \frac{Q^2}{r^2(x)}
		\equiv 1-\frac{2 M}{\sqrt{x^2+a^2}} + \frac{Q^2}{x^2+a^2},     
\eeq
  where $Q$ is the electric charge parameter, such that at $Q=0$ we return to the S-SV metric.    
  At $a=Q=0$, the RN-SV metric reduces to the Schwarzschild metric, while at $a\neq 0$ it describes a regular 
  black hole or a traversable wormhole. 
  
  At $a\neq 0$, the RN-SV metric is manifestly globally regular since all metric 
  coefficients are finite and smooth, hence all its curvature invariants are finite in the whole range $x\in \R$.
  For example, the Ricci scalar $R$, the quadratic Ricci invariant $R\mn R\MN$ and the Kretschmann scalar
  ${\cal K} = R_{\mu\nu\rho\sigma}R^{\mu\nu\rho\sigma}$ have the following values at $x=0$:
\bear
	R  &=&  \frac{2 \left(-3 a M + a^2 + Q^2\right)}{a^4},
\nn
	R_{\mu\nu} R^{\mu\nu} &=&
	\frac{2 \left(2 a^4+a^2 (9 M^2+2 Q^2) - 2 a M (3 a^2+4 Q^2) + 2 Q^4\right)}{a^8},
\nn
	{\cal K} &=&
	\frac{4 \left(3 a^4+a^2 (9 M^2+4 Q^2) - 2 a M (4 a^2+5 Q^2)+3 Q^4\right)}{a^8}.
\ear
  All these expressions diverge as $a\to 0$.
\begin{figure}
	\begin{tabular}{c c c}
		\includegraphics[scale=0.5]{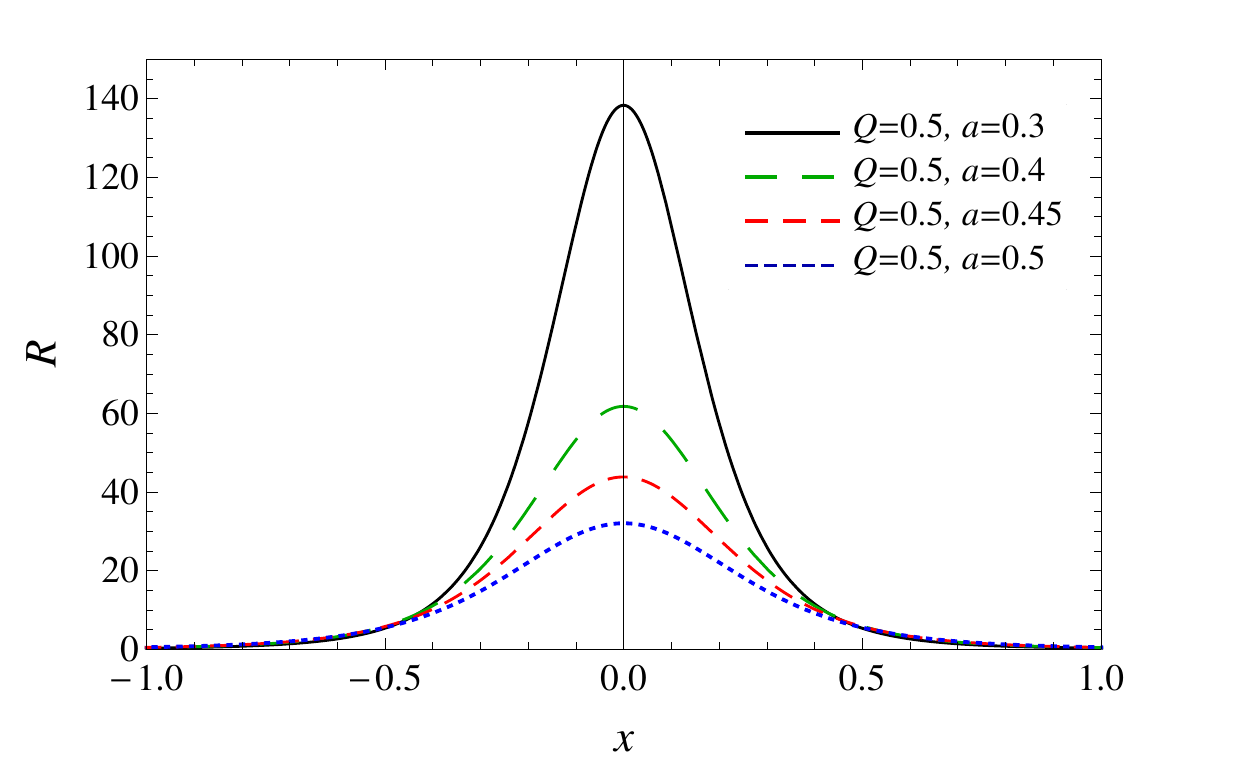}\hspace*{-0.7cm}&
		\includegraphics[scale=0.5]{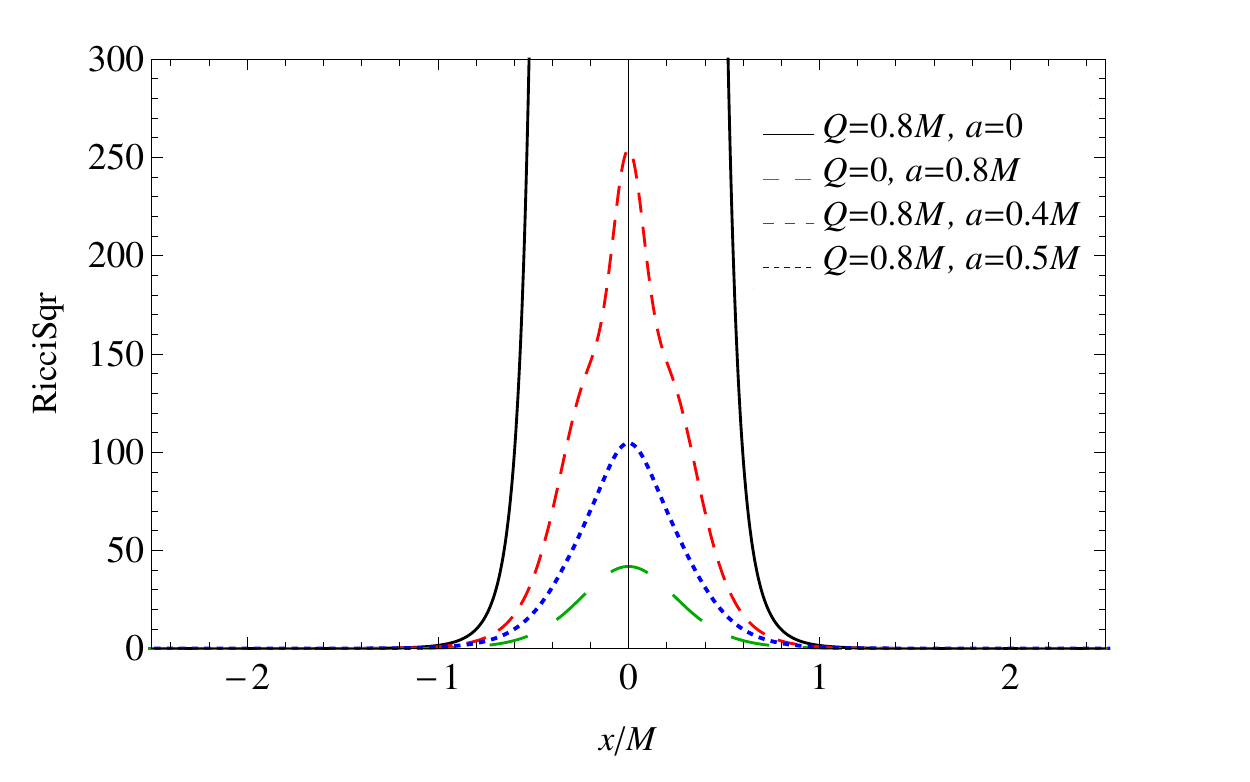}\hspace*{-0.7cm}&
		\includegraphics[scale=0.5]{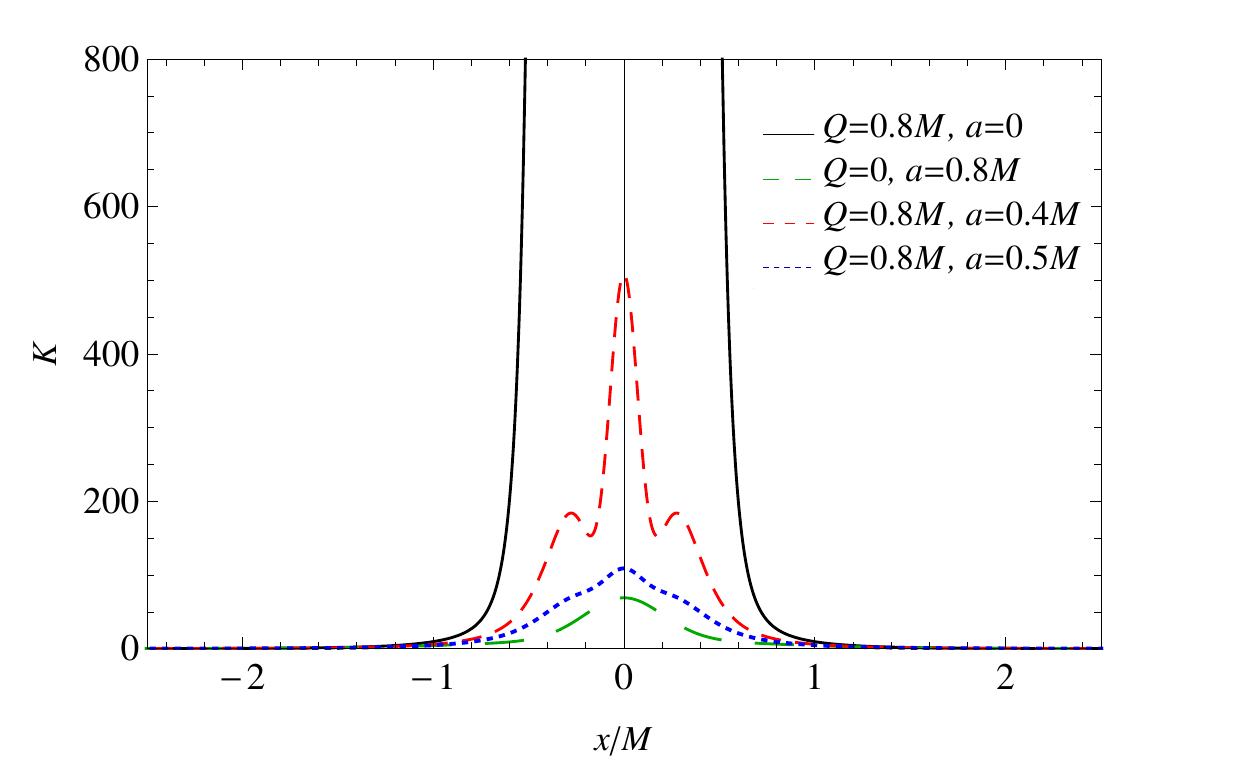}	
	\end{tabular}
\caption{The Ricci scalar, quadratic Ricci invariant and Kretschmann scalar behavior with different 
	$Q$ and $a$. Evidently, the space-time has no singularity if $a\neq 0$.}\label{curvature}
\end{figure}

  The causal structures of RN-SV space-times are more diverse than those of the \RN\ geometries.
  Depending on the values of $a$, $M$ and $Q$, the following kinds of geometry are obtained:
\begin{itemize}
\item 
	a regular black hole with two horizons (curves 1 and 4 in Fig.\,\ref{f1});
\item 
	a regular black hole with a single extremal (double) horizon at $x=0$ (curve 2 in Fig.\,\ref{f1}),
	such geometries exist with both zero and nonzero $Q$;		
\item 
	a regular black hole with two simple horizons and an extremal (double) horizon separating 
	two T-regions (curve 5 in Fig.\,\ref{f1});
\item 	
	a regular black hole with four horizons (curve 6 in Fig.\,\ref{f1});
\item
	a regular black hole with two extremal (double) horizons separating R-regions (curve 7 in Fig.\,\ref{f1});	
\item 
	a traversable wormhole (curves 3 and 8).
\end{itemize}
  One can verify that this list exhausts all possible horizon allocations at $a > 0$, $M >0$ and any values of $Q$.
  The horizons are located at regular zeros of the function $A(x)$, hence at
\beq
		x= x_h =\pm \sqrt{2M^2-a^2-Q^2\pm 2M\sqrt{M^2-Q^2}}.
\eeq   
  
\begin{figure*}[t]
\centering
\includegraphics[width=17cm]{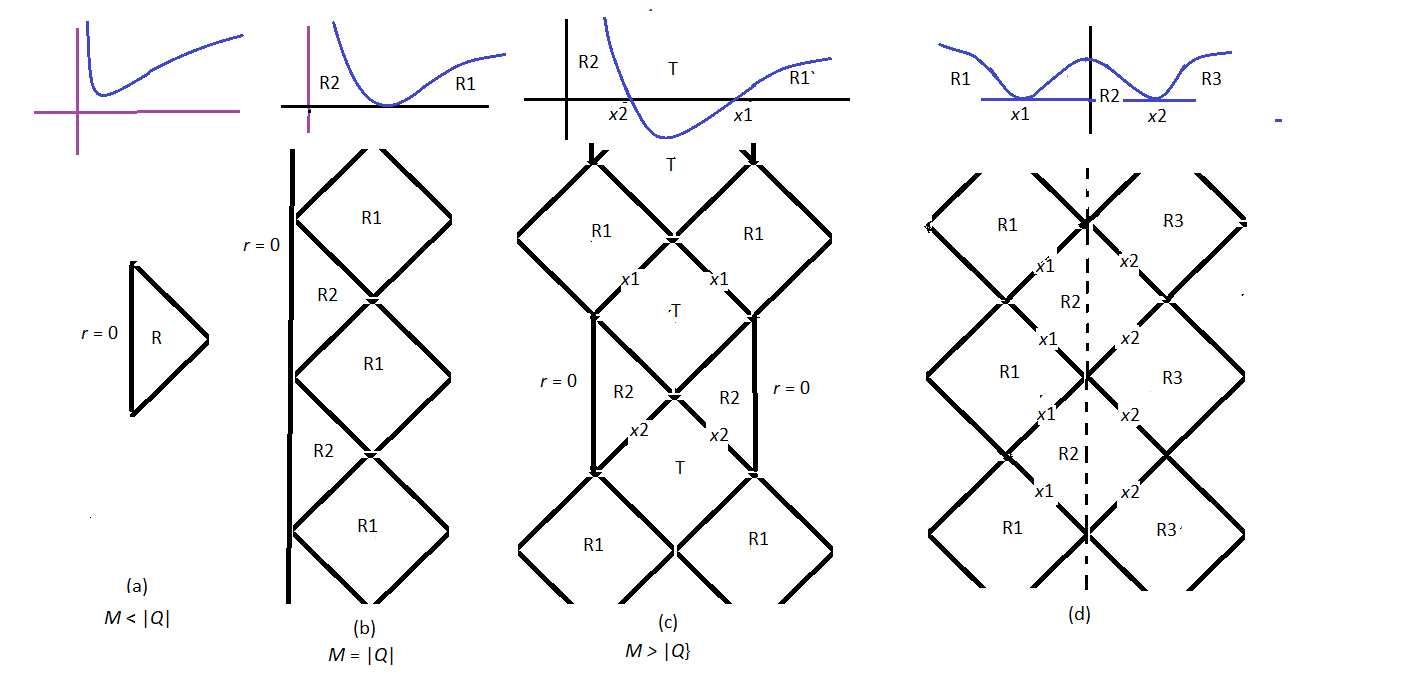}\qquad
\caption{\small  
     Carter-Penrose diagrams for the \RN\ geometries (a, b, c), for comparison, and the RN-SV geometry 
     with two extremal horizons (d). The corresponding behavior of $A(x)$ is shown over the diagrams. 
	All tilted boundary lines in the diagrams correspond to $r = \infty$, all tilted internal lines to horizons.
	The letters R and T mark R- and T-regions, enumerated where necessary, $x1$ and $x2$ mark
	the positions of horizons. The dashed line in (d) shows the throat $x=0$ in the region R2.
	The diagrams (b, c, d) are infinitely continued upward and downward.
	} \label{f3}
\end{figure*}
\begin{figure*}[t]
\centering
\includegraphics[width=17cm]{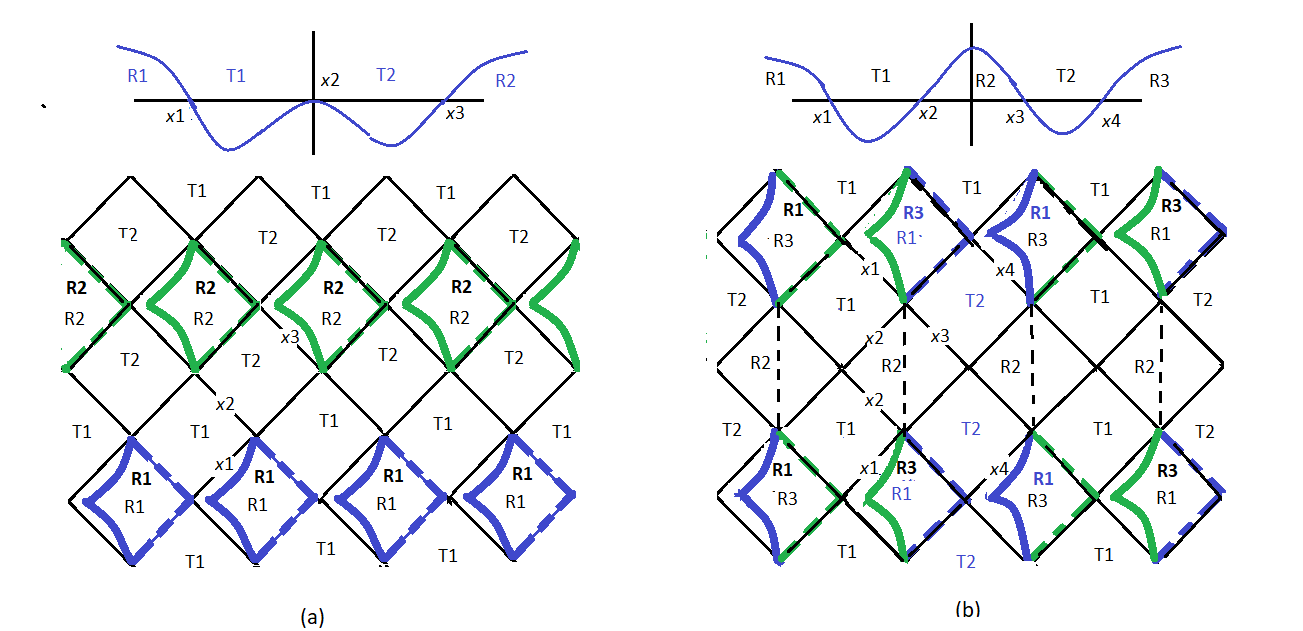}\qquad
\caption{\small  
     Carter-Penrose diagrams for the RN-SV geometries with three (a) and four (b) horizons,
     with the qualitative behavior of $A(x)$ shown at the top. 
	All thick lines in the diagrams correspond to $r = \infty$, all tilted thin lines to horizons.
	Dashed thick lines mean that the corresponding spatial infinity is placed on the lower,
	``directly invisible'' layer of the diagram, under an overlapping horizon. The letters R and T 
	mark R- and T-regions with appropriate numbers. Numbers of the horizons ($x1, x2, \ldots$)
	are easily determined according to regions they separate; some of them are shown as examples.
	The sphere $x=0$ coincides with the horizon $x2$ between the regions T1 and T2 in the left diagram; 
	in the right one, it is a throat in R2, shown by thin dashed lines. Each of these diagrams occupies the whole 
	plane plus a countable set of overlappings.
	}  \label{f4}
\end{figure*}

  As is clear from \rf{ds-gen} with $a>0$ (to be called {\it general SV space-times}), 
  we have everywhere $r \geq a$ with a minimum at $x=0$, which is a throat if $A(0) \geq 0$, 
  a ``black throat'' in the case $A(0)=0$, and a black bounce 
  if $A(0) < 0$. In all cases, space-times with the metric \rf{ds-gen} have no center.\footnote
  		{A center is understood as a location in space where the coordinate spheres shrink to a point, hence, it 
  		 means $r = 0$ while $A > 0$. Note that in a region where $A < 0$, that is, in a T-region, the parameter
  		 $r$ is only time-dependent, and a possible limit $r\to 0$ is actually a cosmological singularity occurring 
	      at a particular time instant. This happens, for example, at the \Scw\ singularity $r\to 0$.}
  Thus regular black hole space-times with the metric \rf{ds-gen} are basically different from the great number 
  of well-known regular black holes which contain either Minkowsky \cite{Simpson:2019mud} or de Sitter 
  regular centers 
  \cite{Bardeen:1968,Hayward:2005gi,AyonBeato:1998ub,Bronnikov:2000yz,kb01-NED,dym92,kb-dym12}, 
  see also references therein. Other regular \sph\ \bhs\ without a center were obtained in \cite{kb-dehn03} as
  examples of vacuum brane-world configurations and in 
  \cite{Bronnikov:2005gm,Bronnikov:2006fu,Bolokhov:2012kn} as black-universe space-times.
  
  Assuming the validity of Einstein's field equations $G\mN = - T\mN$ for SV spacetimes, let us 
  calculate the components of the corresponding stress-energy tensor (SET) (the prime stands for $d/dx$): 
\bearr          \label{T00}
      T^t_t = -G^t_t = \frac{1}{r^2}[1 - A(2rr'' + r'{}^2)) + A' rr']
          = -\frac 1 {r^7} \Big[a^2 \left(r  (2 Q^2+x^2)-4 M x^2\right) - Q^2 x^2 r +a^4 (r -4 M)\Big],
\yyy          \label{T11}
	T^x_x = - G^x_x = -\frac 1 {r^2}[1 - A' r r' - A r'{}^2]
	                         = \frac 1 {r^4} \Big[a^4 + x^2 (a^2+Q^2)\Big],
\yyy	                             \label{T22}      
     T^\theta_\theta = - G^\theta_\theta = - \frac 1r [A r'' + r A''/2 + A'r']
         = - \frac 1{r^7} \Big[a^2 x^2 (r -M) + Q^2 x^2 r + a^4 (r -M)\Big] = T^\varphi_\varphi.
\ear
  Here, we first present the general form of nonzero $G\mN$ for the metric \rf{ds-gen}
  with arbitrary $r(x)$ and $A(x)$, and only then give particular expressions for $T\mN$ with 
  $r \equiv \sqrt{x^2 + a^2}$ and $A(x)$ given by \rf{RN-SV}.
  
  There is an important subtle point about $T\mN$, not always correctly taken into account 
  in descriptions of black-hole studies. The point is that $T^t_t = \rho$, the energy 
  density, and $-T^x_x = P_r$, the radial pressure, only under the condition $A >0$, in other words,
  in static regions (R-regions). In regions where $A < 0$ (T-regions), $x$ is a temporal coordinate, $t$ 
  is a spatial one, therefore, it is $T^x_x$ that should be identified as $\rho$ whereas $T^t_t = -P_r$. 
  
  Concerning the Null Energy Condition (NEC) that in all cases requires $\rho + P_r \geq 0$, 
  we can notice that $\rho + P_r  = (T^t_t - T^x_x) \sign A(x)$. On the other hand, due to the 
  Einstein equations we have the relation 
\beq
		T^t_t - T^x_x = - 2 A r''/r
\eeq  
  for the general metric \rf{ds-gen} with any $A(x)$ and $r(x)$. We see that with both positive
  and negative $A(x)$ it holds $\rho + P_r \propto - r''/r$. But we have necessarily $r'' > 0$ near a 
  minimum of $r$, be it a \wh\ throat or a black bounce. An important conclusion for all \ssph\ 
  space-times is that \cite{BR-book, Lobo:2020ffi}

\medskip  
  {\sl The NEC is necessarily violated in a neighborhood of any \wh\ throat or any black bounce.}  

\medskip
  With $r(x) = \sqrt{x^2 + a^2}$ we have $r''/r = a^2/r^4 >0$, hence the NEC is violated in the whole
  space-time (except for horizons where $T^t_t - T^x_x = 0$ due to $A=0$).
  The Weak Energy Condition (it requires $\rho \geq 0$ in addition to the NEC) is thus also violated.
  The Strong and Dominant Energy Conditions are in such cases most often but not necessarily violated
      
  Another NEC requirement, $\rho +P_\bot \geq 0$ (where $P_\bot = -T^\theta_\theta = -T^\varphi_\varphi$)
  does not lead to equally crucial conditions. For the RN-SV metric we have  
\beq 
     \rho + P_\bot = \frac{3 a^2 M}{r^5} - \frac{2 Q^2 (a^2-x^2)}{r^6}.
\eeq
    
\section{Field sources of the SV metrics}\label{sec-3}

  Let us consider the possible sources of SV geometries in the framework of \GR, in the form of 
  an uncharged scalar field $\phi(x)$ and a nonlinear electromagnetic field with the Lagrangian $\mathcal{L(F)}$,
  minimally coupled to gravity, so that the action may be written in the form
\beq                       \label{Action}
		S=\int\sqrt{-g}d^4x \Big(\mathcal{R} + 2\ep g\MN \d_\mu\phi \d_\nu\phi - 2V(\phi) 
			- \mathcal{L(F)} \Big),
\eeq
  where $\ep = \pm 1$, so that $\ep =+1$ corresponds to a canonical scalar field with positive kinetic energy,
  and $\ep = -1$ to a phantom scalar field with negative kinetic energy; 
  $\mathcal{L(F)}$ is a gauge-invariant NED Lagrangian density with $\mathcal{F} = F_{\mu\nu}F^{\mu\nu}$,
  $F\mn$ being the \emag\ field tensor. Varying the action (\ref{Action}) 
  with respect to the metric $g^{\mu\nu}$ leads to the Einstein equations
\beq                     \label{FieldEq}
               G\mN = - T\mN[\phi] - T\mN[F],
\eeq
  where $T\mN[\phi]$ and $T\mN[F]$ are the SETs of the scalar and electromagnetic fields:
\bearr                   \label{SET-phi}
			T\mN[\phi] = 2\ep \d_{\mu}\phi\d^{\nu}\phi 
				- \delta\mN \left(\ep g^{\rho\sigma}\d_\rho \phi \d_\sigma\phi -V(\phi)\right),
\yyy                    \label{SET-F}
			T\mN[F] = - 2 \mathcal{L_F} F_{\mu\sigma} F^{\nu\sigma} 
					+\frac 12 \delta\mN \mathcal{L(F)},
\ear 
  where $\mathcal{L_F} = d\mathcal{L}/d\mathcal{F}$; 
  further on, varying the action in $\phi$ and $F\mn$, we obtain the field equations
\bearr         \label{eq-phi}
			2 \ep \nabla_{\mu}\nabla^{\mu}\phi + d V(\phi)/d \phi =0,
\yyy		\label{eq-F}
			\nabla_\mu(\mathcal{L_F}F\MN) = 0.
\ear

  Considering the \ssph\ metric \rf{ds-gen}, the corresponding assumptions for the scalar and \emag\ fields
  are $\phi = \phi(x)$ and possible nonzero $F\mn$ components: $F_{tx}= -F_{xt}$ (a radial electric field) 
  and $F_{\theta\varphi} =-F_{\varphi\theta}$ (a radial magnetic field). We will assume the existence 
  of a magnetic field only, such that $F_{\theta\varphi} = q \sin\theta$, where $q$ is the magnetic 
  monopole charge. Equation \rf{eq-F} is then trivially satisfied. The Faraday \emag\ invariant takes the form 
  $\mathcal{F} = 2 q^2/r^4 \equiv 2q^2/(a^2+x^2)^2$. Under these assumptions, the SETs  
  \rf{SET-phi} and \rf{SET-F} take the form
\bearr         \label{T-phi}
		T\mN[\phi] = \ep A(x) \phi'{}^2 \diag (1, -1, 1, 1) + \delta \mN V(\phi), 
\yyy           \label{T-F}		
		T\mN[F] = \frac 12 \diag\Big(\mathcal{L},\ \mathcal{L},\
				   \mathcal{L} - \frac{4q^2}{r^4}\mathcal{L_F},\ 
				   \mathcal{L} - \frac{4q^2}{r^4}\mathcal{L_F},\Big). 
\ear  
  
  Let us now explain why a scalar field alone cannot be a source of SV metrics. To begin with, according to
  the most universal of the so-called global structure theorems \cite{BR-book, kb01-glob}, an \asflat\ \ssph\
  Einstein-scalar configuration cannot contain more than one horizon, while the SV metrics with proper
  parameters contain at least two. But more specifically, the scalar field SET has the property
  $T_t^t[\phi] = T^\theta_\theta[\phi]$, and this is clearly not the case for \rf{T00} and \rf{T22},
  the SET components needed for SV metrics.
  
  On the other hand, NED alone also cannot provide a source for SV metrics since, for the latter,
  the SET components \rf{T00} and \rf{T11} are not equal, whereas by \eqn{T-F}, $T^t_t [F] = T^x_x [F]$.
  Moreover, an attempt to use a scalar field combined with a Maxwell field (which is a special case of NED
  corresponding to  $\mathcal{L = F}$) would also be a failure since for a Maxwell field 
  $T_t^t - T^\theta_\theta = 4q^2/r^4$, while the corresponding difference for the SV metrics 
  exhibits another $x$-dependence. 
  
  It can, however, be shown that a combination of a scalar field and NED can provide a source for any SV
  metric \rf{ds-gen} with an arbitrary function $A(x)$ and $r = \sqrt{x^2+a^2}$. 
  
  Indeed, the difference of \eqs \rf{T00} and \rf{T22} is free from the scalar field and leads to
\beq
  		\frac 1 {r^4}[2 - 2 A(x) + r^2 A''(x)] = \frac{2q^2}{r^4} \mathcal{L_F}.
\eeq    
  The left-hand side of this equation is, by our assumptions, a known function of $x$, and since we have
  $\mathcal{F} = 2 q^2/r^4 \equiv 2q^2/(a^2+x^2)^2$, it is easy to express
  $d\mathcal{L}/dx = \mathcal{L_F}d\mathcal{F}/dx$ in terms of $x$. Integrating, we then 
  obtain $\mathcal{F}$ as a function of $x$ and consequently of $\mathcal{F}$. 
   
  Next, the difference of \rf{T00} and \rf{T11} is free from a NED contribution and leads to 
\beq
			\phi'{}^2 = - \ep \frac{a^2}{x^2+a^2} = \frac{a^2}{x^2+a^2} \ \then\ 
			\phi = \pm \arctan \frac xa + \phi_0, \qquad \phi_0 = \const.
\eeq    
  It is evident here that the scalar field should be phantom, $\ep = -1$, and without loss of generality
  we can put $\phi_0 = 0$.
  
  Then, using the already known expressions for $\mathcal{L}$ and $\phi'{}^2$ as functions of $x$,
  we can use \eqn{T11} to calculate the potential $V$ as a function of $x$ and consequently of $\phi$:
\beq
			V(x) = - A\phi'{}^2 - \frac 12 \mathcal{L} - G^x_x.
\eeq   
  This completes the calculation. The scalar field equation \rf{eq-phi} is a consequence of the Einstein 
  equations, and it is not necessary to consider it, but it can be used in order to verify the correctness 
  of the results.    
  
  Using this algorithm for the function $A(x)$ given by \rf{RN-SV} (the RN-SV metric), we obtain 
\bearr              \label{L-fin}
          \mathcal{L(F)} = \frac{12}{5}\,\frac{Ma^2}{(x^2+a^2)^{5/2}} 
          					+ \frac {2Q^2 (3x^2-a^2)}{3(x^2+a^2)^3}
          		= \frac{12 Ma^2}{5 (2q^2/\mathcal{F})^{5/4}} 
			+ \frac {2Q^2 \big[3(2q^2/\mathcal{F})^{1/2}-4a^2\big]}{3(2q^2/\mathcal{F})^{3/2}},			
\yyy                \label{V-fin}        					 
            V(\phi) = \frac {2a^2 [6M \sqrt{x^2+a^2} - 5Q^2]}{15 (x^2+a^2)^3}
                 = \frac {2 \cos^6 \phi}{15 a^4} (6M a \sec \phi - 5Q^2).
\ear   
\begin{figure}
	\begin{tabular}{c c}
		\includegraphics[scale=0.7]{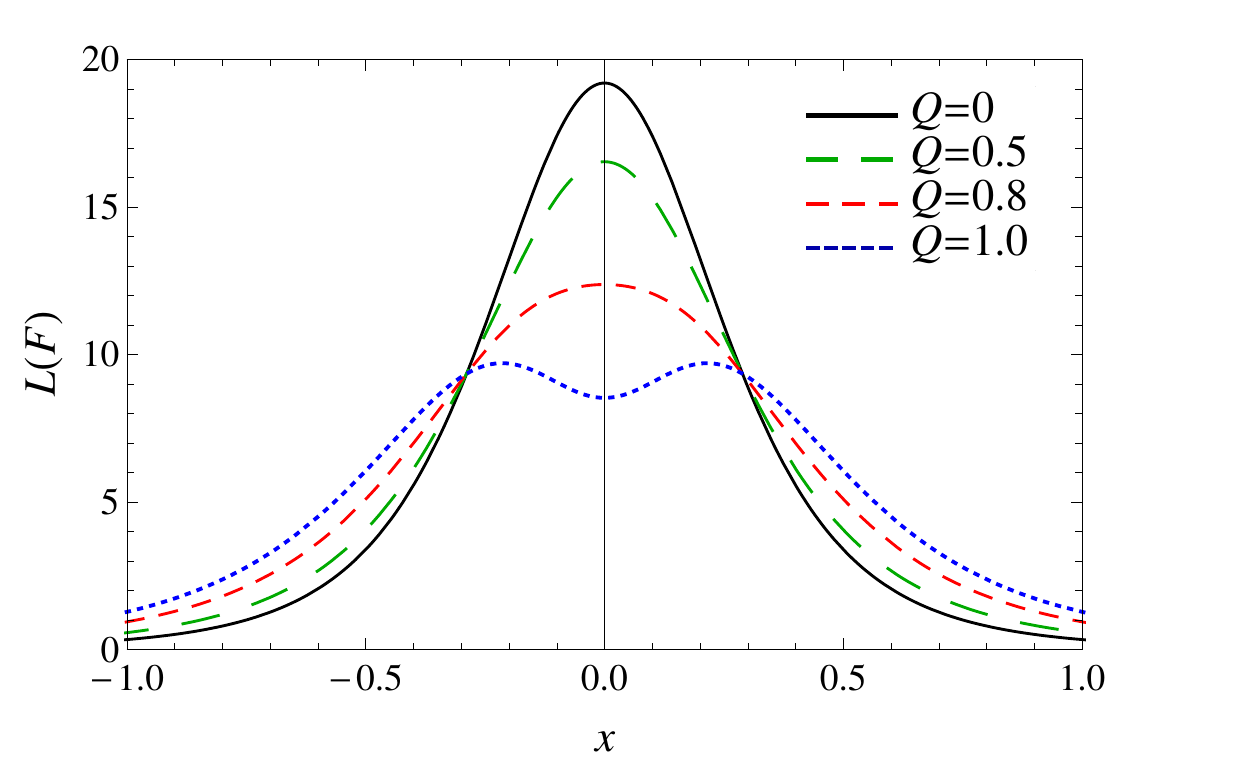}\hspace{-2cm}&
		\includegraphics[scale=0.7]{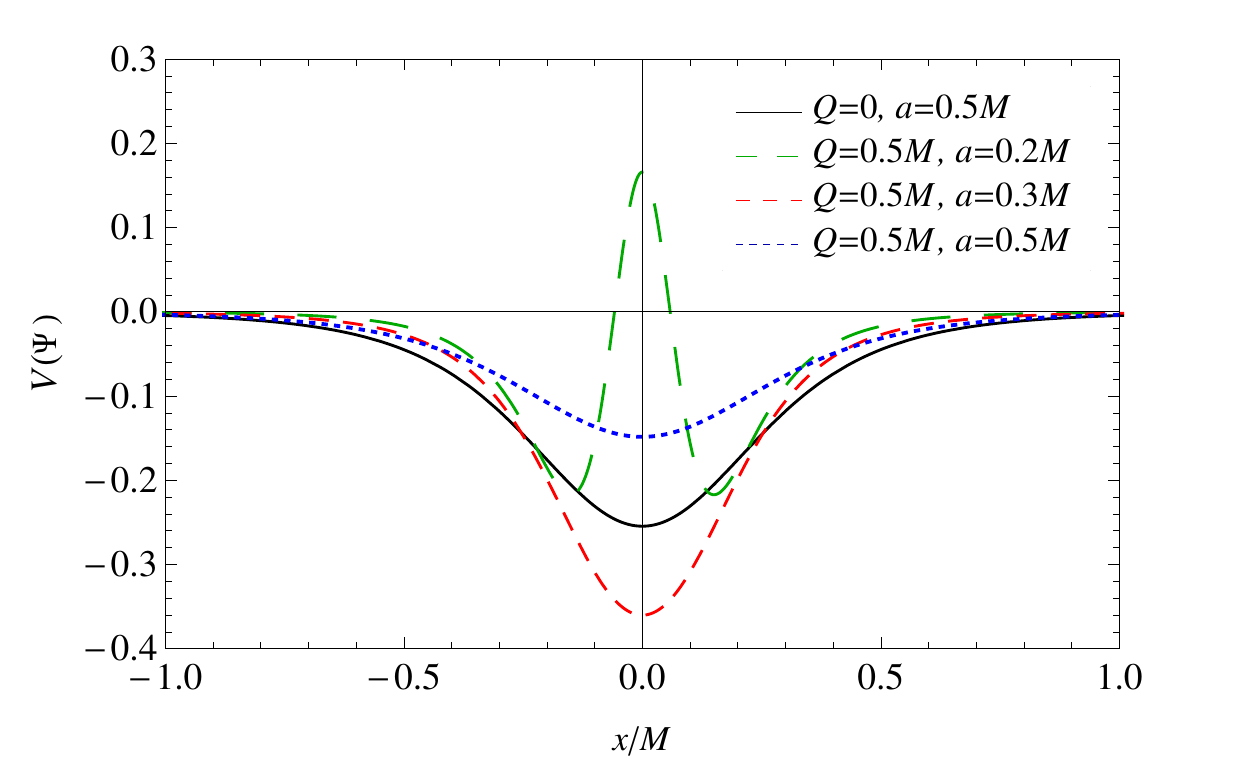}
	\end{tabular}
	\caption{The NED Lagrangian density $\mathcal{L(F)}$ and the field potential $V(\phi)$ 
	behavior with different $Q$ and $a$.}               \label{SET}
\end{figure}

  At $Q=0$, we obtain the corresponding results for the S-SV space-time. 
  
  We notice a curious feature of the sources for RN-SV geometry. The metric itself contains the charge 
  parameter $Q$, associated in the original \RN\ metric with an electric or magnetic charge in Maxwell's 
  electrodynamics. We have found, however, that our source of the RN-SV solution does not contain any
  Maxwell field but, instead, is sourced by a magnetic field obeying NED; moreover, according to \eqn{L-fin},
  for a given value of the parameter $Q$ in the metric, the source also contains a separate free parameter $q$,
  the NED magnetic charge, and the corresponding Lagrangian function $\mathcal{L(F)}$ depends on both. 
  In Fig.~\ref{SET}, the behavior of NED field Lagrangian density and the scalar field potential is shown 
  for different $Q$ and $a$. As expected, $\mathcal{L(F)}$ is finite everywhere.
  
  A certain shortcoming of this construction is that the function $\mathcal{L(F)}$ in \rf{L-fin} contains fractional 
  powers of $\mathcal{F}$, which makes meaningless negative values of $\mathcal{F}$. This, in particular, 
  explains why we use magnetic fields with $\mathcal{F}> 0$ rather than electric ones implying $\mathcal{F}< 0$. 
  An attempt to create a similar construction with electric fields also leads to fractional powers of $\mathcal{F}$. 
  If we try to employ $|\mathcal{F}|$ instead of $\mathcal{F}$, the function $\mathcal{L(F)}$ still loses 
  analyticity at $\mathcal{F}=0$.

\section{Concluding remarks}\label{sec-5}

  Simpson and Visser \cite{Simpson:2018tsi} have considered a very simple and theoretically appealing 
  spherically symmetric black bounce space-time (SV space-time), which modifies the Schwarzschild black hole 
  metric in a precisely controlled and minimal manner and interpolates between regular black holes and 
  traversable wormholes. The new length scale parameter $a$ is responsible for the curvature singularity 
  regularization and can potentially mimic the quantum gravity effects at a small-length scale. Further on, 
  the charged (RN-SV) and rotating SV spacetimes have also been proposed. All of them are globally regular,
  and in particular, in S-SV and RN-SV regular black holes, their $r=0$ singularities are replaced by a black 
  bounce that ultimately leads to another \asflat\ universe. This class of space-times generalizes and broadens 
  the well-known variety of regular black holes and seems to be very intriguing owing to its simplicity and a unified 
  treatment of distinct kinds of geometries. 
  
  As we saw in the present paper, these space-times have in general multiple horizons and complex 
  global causal structures. Depending on the values of $a$, the $x=0$ hypersurface can have a different nature:
  a wormhole throat, a black bounce, or a black throat in the intermediate case. The RN-SV space-time as a 
  whole describes a variety of geometries including traversable wormholes and different kinds of regular black holes.

  With the radius $r(x)$ having a minimum, the corresponding SET, as expected, violates the NEC and other 
  energy conditions. While canonical scalar fields cannot lead to wormhole or regular black hole geometries 
  \cite{kb01-glob}, it is possible with phantom scalar fields \cite{Bronnikov:2005gm,Bolokhov:2012kn}. 
  Still we have seen here that SV-like geometries cannot be sourced by a scalar field or NED taken separately.
  However, we have shown that the RN-SV and S-SV spacetimes are exact solutions of the Einstein field equations
  with a combination of these two sources. This work is important in the light that it uplifts the status of 
  SV-like space-times from ad-hoc mathematical models to a class of exact solution of gravitational field equations. 

  Let us also stress that the algorithm we have used makes it possible to find a similar combination of field sources
  for any \ssph\ SV geometry with the metric \rf{ds-gen}. These can include, for example, the proper modifications 
  of \RN-de Sitter space-times, with $A(x) = 1 -2M/r + Q^2/r^2 - \Lambda r^2/3$,
  which will then possess up to six horizons and accordingly very complex causal structures. 
  The sources of black bounce space-times proposed in \cite{Lobo:2020ffi} can also 
  be easily calculated. A task of interest for a future study can be to find a source for SV-like geometries with rotation 
  \cite{Mazza:2021rgq, Xu:2021lff, Shaikh:2021yux, SV21-kerr}. 
  
  Another issue of importance is the stability of SV space-times, which problem cannot be formulated until
  we know their precise dynamic context. In this regard, it can also be of interest to find possible alternative 
  sources of the same SV space-times, because the stability properties of the same geometry may be different 
  in the presence of different sources of this geometry. This problem was, in particular, discussed in 
  \cite{kb-sha13, kb-fa21} using as an example the simplest Ellis wormhole \cite{Ellis, kb73}.  

\subsection*{Acknowledgment}

 The work of K.B. is supported in part by the RUDN University Strategic Academic Leadership Program, by RFBR
 Project 19-02-00346, and by the Ministry of Science and Higher Education of the Russian Federation, Project
 ''Fundamental properties of elementary particles and cosmology" N 0723-2020-0041. 
 R.K.W. thanks the NRF and the University of KwaZulu-Natal for continued support. R.K.W. is also grateful 
 to Prof.~Sunil Maharaj, Prof. Sushant Ghosh and Prof.~Rituparno Goswami for fruitful discussions.


\end{document}